\documentclass[conference]{IEEEtran}
\IEEEoverridecommandlockouts
\ifCLASSINFOpdf
\else
\fi
\usepackage{amsmath, amssymb}
\usepackage{enumitem, array}
\usepackage{tabularx}
\usepackage{graphicx}
\usepackage{textcomp}
\usepackage{xcolor}
\usepackage{color}
\usepackage{booktabs}

\newcommand{\MH}{\color{black}}

\begin{document}
%
\title{WIP: Energy-Efficient LLM-based Serving Cluster Formulation in Cell-Free Massive MIMO
\thanks{M. Hoffmann was funded by the Polish National Science Centre, project no. 2022/45/N/ST7/01930. P. Kryszkiewicz was funded by the Polish National Science Centre, project no. 2021/41/B/ST7/00136 and 
2023/05/Y/ST7/00002 (PASSIONATE project within the European Coordinated Research
on Long-term Challenges in Information and Communication Sciences and
Technologies CHIST-ERA Programme).}
}

\author{
\IEEEauthorblockN{Marcin Hoffmann}
\IEEEauthorblockA{\textit{Poznan University of Technology} \\
}
\and
\IEEEauthorblockN{Pawel Kryszkiewicz}
\IEEEauthorblockA{\textit{Poznan University of Technology} \\
}
}


\maketitle

\begin{abstract}
One way to increase the Energy Efficiency (EE) of 6G wireless networks is to utilize existing network infrastructure more efficiently. This can be achieved by introducing User-Centric Cell-Free Massive Multiple-Input-Multiple-Output (UCCF MMMIMO), which allows for simultaneously serving a single user by multiple Base Stations (BSs). From this perspective, the key challenge is to decide which BSs should serve a given user, known as the Serving Cluster Formulation (SCF). In this paper, we propose to deal with this problem by using an Artificial Intelligence (AI) agent based on a Large Language Model (LLM), targeting improvement of EE.
We evaluated the proposed AI agent using a complex, 3D Ray Tracer-based, cellular network simulator, 
comparing a few GPT models and state-of-the-art algorithms. The results show up to 32\% gain in EE of the proposed AI agent compared to the baseline.
\end{abstract}


%
\IEEEpeerreviewmaketitle

\section{Introduction}

One of the key technologies for {\MH energy-efficient} 6G wireless networks is the User-Centric Cell-Free Massive Multiple-Input-Multiple-Output (UCCF MMMIMO)~\cite{ammar2022}. In the state-of-the-art network-centric approach, where User Equipment (UE) is typically served by a single Base Station (BS) providing the highest received power. {\MH In this case, some BSs might be underutilized yet consuming a significant amount of energy, leading to reduced network Energy Efficiency (EE).} In contrast, in the UCCF MMIMO network, multiple BSs (sometimes called Access Points) can jointly serve one UE to provide a uniform Quality of Service (QoS) in the network~\cite{buzzi2017}. {\MH As a result, the network offers higher throughput to the users, while utilizing the same hardware resources, providing EE gains.} One of the key challenges for the 6G UCCF MMIMO network is to decide which BSs should serve a particular UE. This is known in the literature as Serving Cluster Formulation (SCF)~\cite{kim2022}.
{\MH Energy-efficient} SCF requires balancing properly between improved UE received power, interference coordination related to serving multiple UEs, and the risk of inefficient allocation of power for transmission to UEs characterized by a significant path-loss for certain BSs. 
Although some authors provide analytical solutions to the SCF in UCCF MMIMO network, these are usually obtained under the simplified system model, e.g., authors of~\cite{Hao2024} propose a joint user association and power control for UCCF MMIMO network, but rely on the single-carrier system, the Rayleigh radio channel model, do not consider space-time-frequency Radio Resource Scheduler (RRS), and neglect nonlinear characteristics of the radio front-end. In contrast, 5G and beyond networks are multi-carrier Orthogonal Frequency Division Multiple Access (OFDMA) systems, with time-frequency-space RRS. Moreover, the realistic radio channels result in non-trivial spatial distortion characteristics affecting the reception of signals in an MMIMO system~\cite{hoffmann2024}. Under the realistic system model, one approach is to utilize a low complexity heuristic algorithm, e.g.,~\cite{cfmimobook2021} proposes to serve UE by the BS being responsible for 95\% of the received power, or the authors of~\cite{D’Andrea2020} suggest serving UEs by a fixed number of the closest BS, indicating 20 BSs as optimal, based on their simulations. However, these {\MH propositions do not consider EE optimization, and} have limited performance because no feedback from the network is taken into account. In this context, a reasonable approach is to utilize Artificial Intelligence (AI), which could infer the proper serving cluster based on the network data. Especially, Reinforcement Learning (RL) enables interaction of the so-called agent with the environment to learn how to act (take action) under certain conditions (state). The RL approach to the SCF was demonstrated in~\cite{Tsukamoto2025}. However, the proposed AI model has a fixed input size, making it difficult to generalize.
Moreover, {\MH EE optimization is not taken into account, and} the utilized system model does not consider OFDMA or a nonlinear radio front end. But the key drawback of the pure RL algorithms is that the agent does not have any embedded intelligence or human-like reasoning and initially treats each action as equally good. This makes the state-of-the-art RL agent learning time significant, and a fixed action space prevents exploring actions of potentially high reward. Recently, the development of Large Language Models (LLMs) has created an opportunity to improve the RL by deploying the AI agent, which, given its role and context, analyzes the state and past experience, taking intent-driven actions. 
An example of such an AI agent based on LLM was demonstrated by the authors of \cite{Bao2025LLMhRICLH}, but did not touch on the UCCF MMIMO network. 

Therefore, in this paper, we propose an LLM-based AI agent whose goal is to perform SCF in the UCCF MMIMO network. The proposed AI agent aims to improve {\MH network EE} based on autonomous interaction with the environment. 
In contrast to state-of-the-art works, to evaluate the proposed solution, we utilize a realistic system-level simulator of the UCCF-MMIMO network based on the 3D Ray Tracer, which we proposed in~\cite {hoffmann2024}. The simulator is considering, e.g., OFDMA, a dedicated space-time-frequency RRS, or nonlinear characteristics of the radio front-end. Simulation studies compare the proposed LLM-based AI agent with state-of-the-art algorithms, while using 3 models: GPT-5, GPT-4o, and GPT-4o-mini.

The rest of the paper is organized as follows. In Sec.~\ref{sec:aiagent} we described the proposed AI agent. Sec.~\ref{sec:results} covers the results of the simulation studies. The paper is concluded in Sec.~\ref{sec:conclusions}.

\section{Energy-Efficient, LLM-based SCF} \label{sec:aiagent}
\begin{figure}[!t]
\centering
\includegraphics[trim={9cm 1.0cm 9cm 1.0cm}, clip, width=0.48\textwidth]{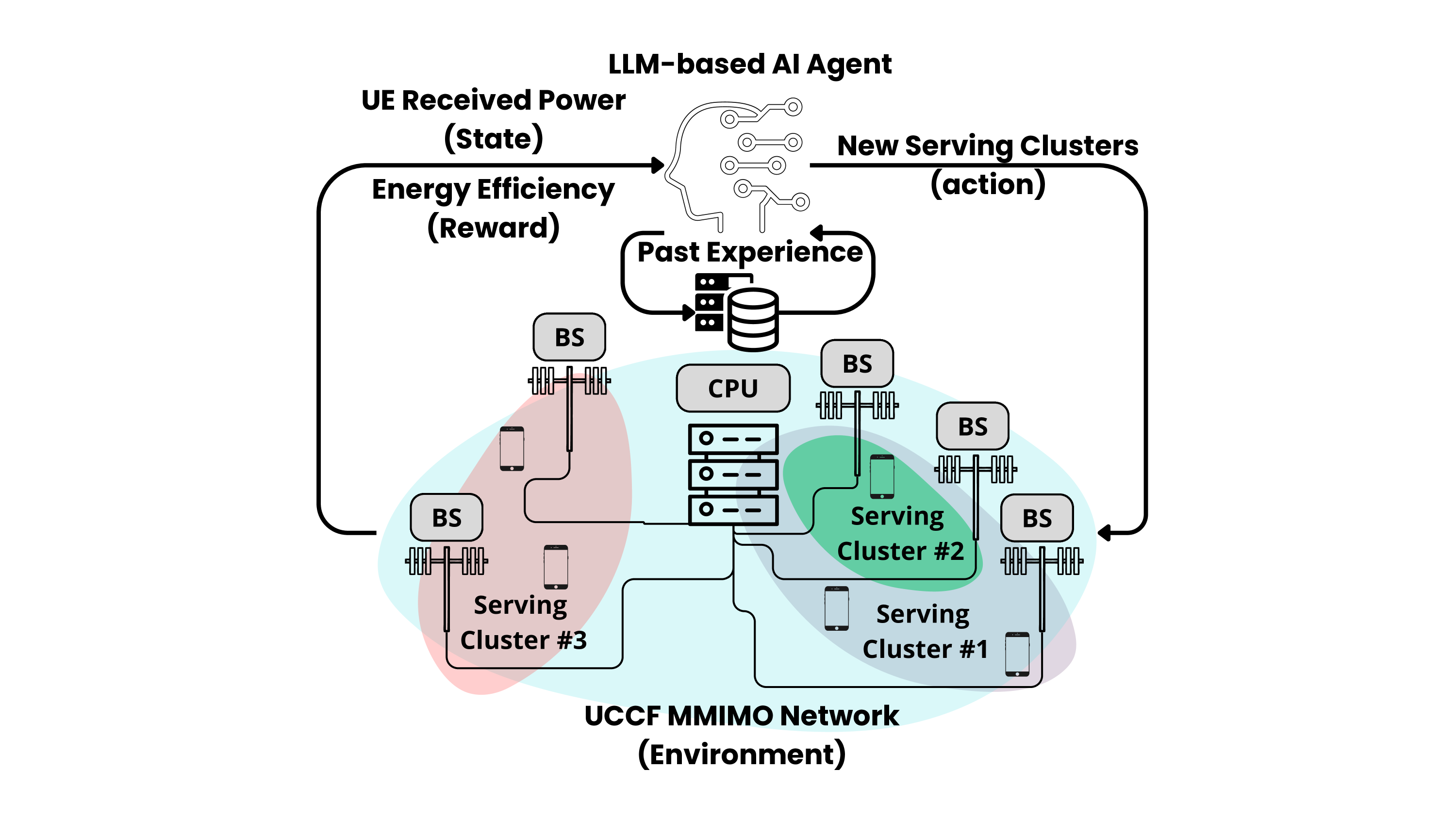}
\caption{Energy-efficient LLM-based SCF in UCCF MMIMO network.}
\label{fig:aiagent}
\end{figure}

The proposed LLM-based AI agent aiming to improve {\MH EE} in the UCCF MMIMO network by dynamic SCF is depicted in~Fig.~\ref{fig:aiagent}. The considered network consists of the so-called Central Processing Unit (CPU) with multiple MMIMO BSs attached. The CPU is responsible for the coordination of the transmission from the BSs, RRS, and high physical layer functions, but most importantly, {\MH hosts the LLM-based AI agent, which performs energy-efficient SCF}. On the other hand, each BS is responsible for the low physical layer functions, e.g., application of precoder weights. This functionally disaggregated architecture fits well with the Open Radio Access Network (RAN) concept, as a candidate for its practical deployment~\cite{hoffmann2024infocom}. 

To perform the SCF, the proposed LLM-based AI agent operates in an RL cycle: it recognizes the state, utilizes past experience to take action, observes the reward, and recognizes a new state, thereby continuing the cycle. In the case of the SCF in the considered UCCF MMIMO network, we propose the following definitions:
\begin{itemize}
    \item \textbf{State} provides an AI agent with information about the powers received in the DL from each BSs by each UE. This is shaped in a matrix form, where each row corresponds to a UE, and each column contains the power received by this UE from a specific BS. This enables the AI agent to recognize which UEs should be served by a single BS, and which by many BSs constituting a serving cluster. 
    \item \textbf{Action} is returned by the AI agent in the form of a vector of length equal to the number of UEs, filled with integers indicating the number of BSs providing the highest received power that should serve a particular UE. 
    \item \textbf{Reward} is defined as {\MH network EE. While the state-of-the-art definition of EE is the total system throughput divided by the total power consumption, we adapt the definition proposed in~\cite{Hoffmann2021}: median user throughput divided by the total power consumption.} Based on the results in~\cite{Hoffmann2021}, covering EE in MMIMO networks, it allows us to maintain more fairness on the users associated with worse radio conditions.
\end{itemize}
The state-of-the-art RL solutions deal with action selection, using algorithms like epsilon-greedy or stochastic policies. In contrast, LLMs offer an intent-driven approach, where the behavior of the AI agent can be defined in terms of having a human-readable text form instead of a strict mathematical algorithm~\cite{Habib2025,Bao2025LLMhRICLH}. From this perspective, the definition of the AI agent requires formulating the so-called \emph{system prompt}, which provides the AI agent with the context of its role and general instructions on how to behave, and \emph{user prompt}, which is the direct command sent to the AI agent to receive output. In the following sections, we present our propositions for \emph{system prompt} and \emph{user prompt}, which we used to build an AI agent aimed at SCF, with a focus on improving {\MH network EE}. For the prompts demonstration purposes, we assumed 3 base stations and 5 UEs, and random values of experience samples. However, the proposed AI agent can deal with any number of UEs and BSs, as will be shown in Sec.~\ref{sec:results}.
\subsection{System Prompt}
        \noindent\textit{You are a mobile network operator AI that controls a wireless 6G network with 3 base stations following the user-centric cell-free paradigm. Users can be served simultaneously by multiple coordinated base stations. It is good to serve the user 
        by multiple base stations if their received power is similar. If the power received by a given user from one base 
        station is significantly higher compared to other ones, it is better to serve this user with only 1 base station.
        \textbf{Your goal is to maximize the \MH{EE} of the network.} I will provide you with your past experience and a 2D array of received power in dBm between each user and each base station (rows = user ID, columns = BS ID). You must return a list of the number of base stations that should serve the given user (ints, 0-based index of user).
}
        \begin{itemize} 
        \item[-] \textit{When there is no past experience, start by assigning each user 1 base station.}
        \item[-] \textit{When there exists past experience, try to increase the number of base stations for certain users.}
        \item[-]\textit{When in the experience increasing the number of base stations improves {\MH EE}, try to increase the number of base stations further for certain users.}
        \item[-] \textit{Every user from the input must appear in the output list.}
        \item[-] \textit{Analyze the received power for each row (user ID) separately.}
        \item[-] \textit{Try to explore new assignment options.}
        \item[-] \textit{The length of the output list must be equal to the number of users (rows) in the input.}
        \item[-] \textit{Use information provided as past experience.}
        \item[-] \textit{when the{\MH EE} from the experience is similar over the latest 3 attempts, try to randomly apply minor modifications.}
        \item[-] \textit{after having more than 40 experience samples, stick to a list of the number of base stations from your experience, which should serve the given user which provided the highest {\MH EE}.} 
        \item[-] \textit{Return ONLY the list with no explanations.}
        \end{itemize}
        
\subsection{User Prompt}
\noindent \textit{Here is your past experience:}
\begin{itemize}
    \item \textit{\textbf{Experience sample 1:} (representative example)}
        \begin{itemize}
            \item \textit{List of number of base stations serving each user: $[1, 1, 1, 2, 1]$}
            \item \textit{List of throughput in Mbps for each user (float, 0-based index of user): $[1, 1.5, 0.5, 2, 1]$}
            \item \textit{{\MH Network EE in kbit/J}: 1}
        \end{itemize}
\end{itemize}

\noindent\textit{Here is the current 2D array of received power in dBm between each user and each base station: [[-80,-75,-60],[-58,-59,-60],[-75,-72,-76],[-65,-66,-80],[-55,-70,-78]]. Based on the past experience and the provided array of received power, decide how many base stations should serve each user to maximize {\MH network EE}.}
\begin{itemize}
    \item [-] \textit{Return ONLY the list as described.}
    \item [-] \textit{Every user from the input must appear in the output.}
    \item[-] \textit{Every user can be served a maximum of 3 base stations.}
    \item [-] \textit{The size of output must be equal to 5.}
\end{itemize}

\section{Simulation Studies} \label{sec:results}

\begin{figure}[!t]
\centering
\includegraphics[trim={0cm 0.0cm 0cm 0cm}, clip, width=0.47\textwidth]{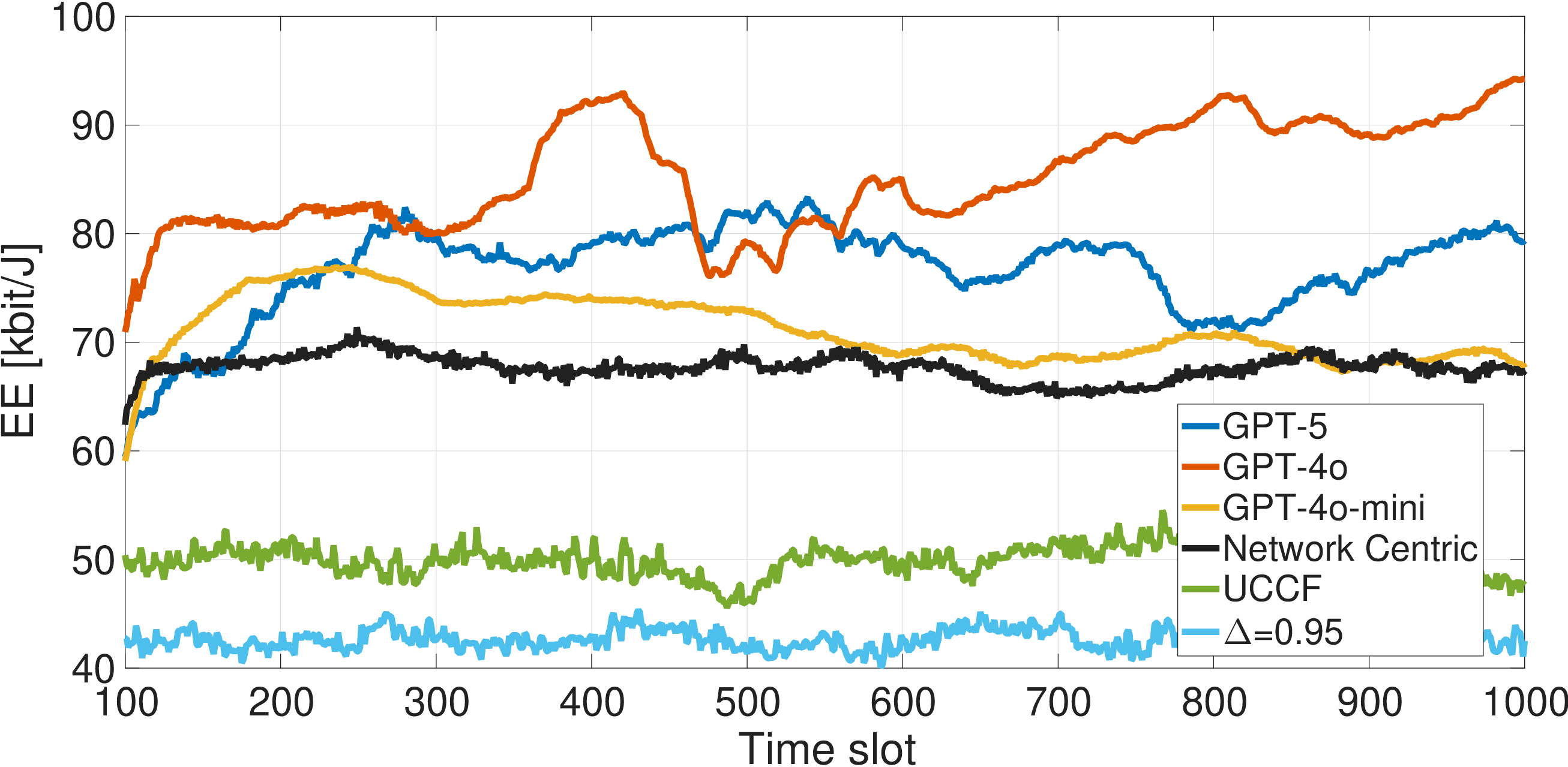}
\caption{EE over time with moving average of 100 slots.}
\label{fig:training}
\end{figure}

\begin{figure}[!b]
\centering
\includegraphics[trim={0cm 0cm 0cm 0cm}, clip, width=0.47\textwidth]{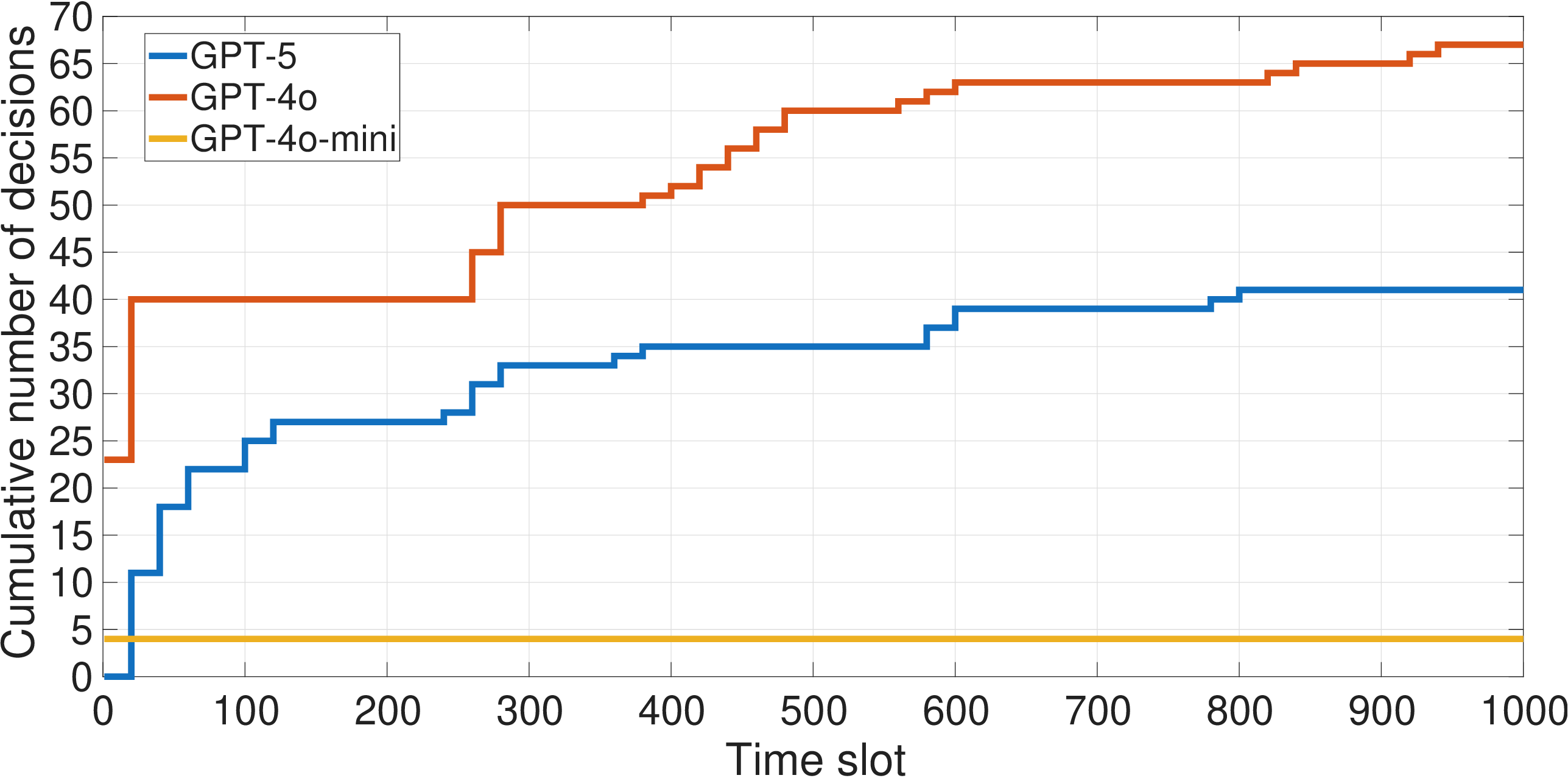}
\caption{Cumulative number of decisions to modify serving cluster for a UE over time, for each considered GPT model.}  
\label{fig:decisions}
\end{figure}

\begin{figure}[!t]
\centering
\includegraphics[trim={3cm 1.0cm 3cm 1cm}, clip, width=0.48\textwidth]{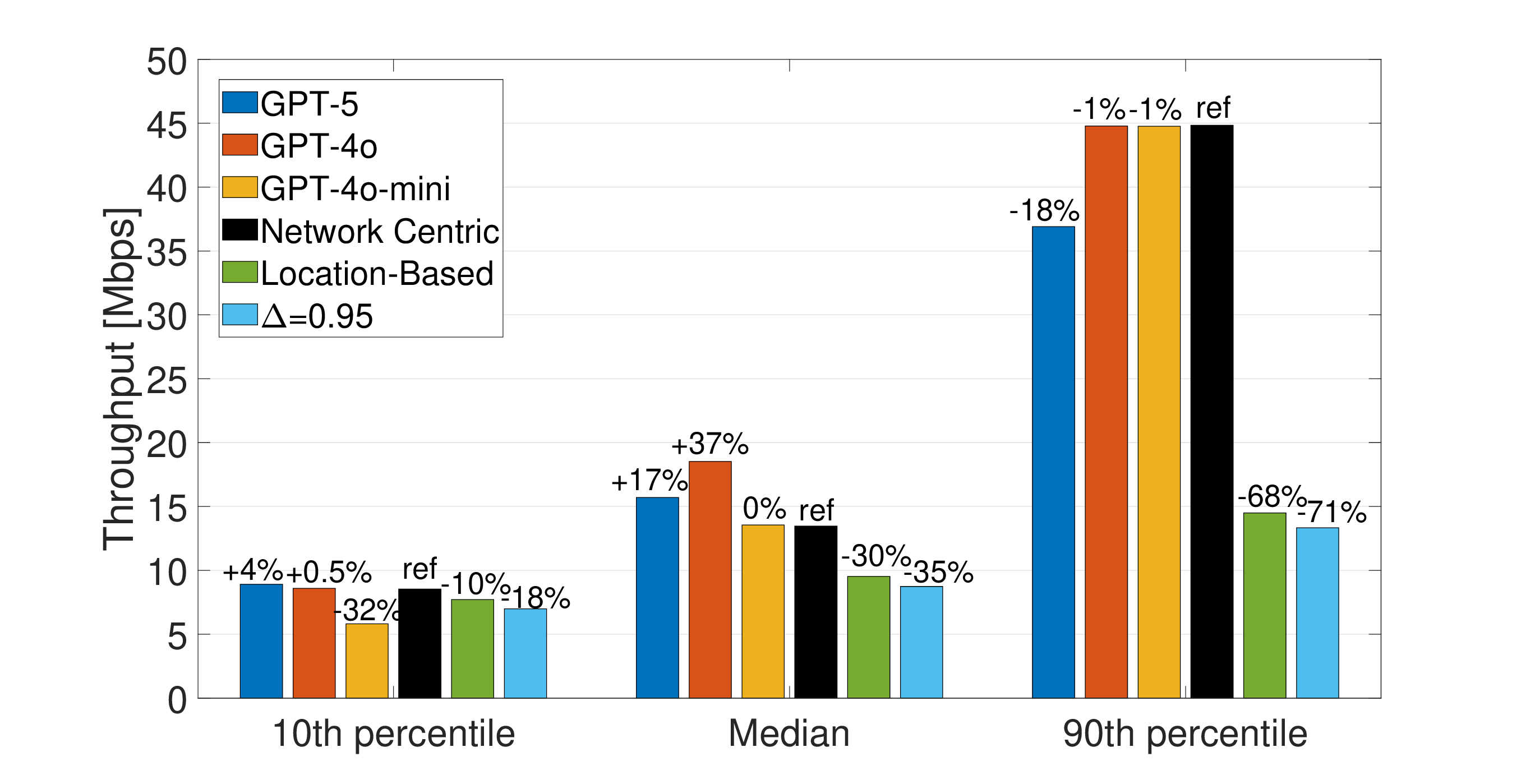}
\caption{10th percentile, Median, and 90th percentile of UE throughput, calculated from the throughput of UEs averaged over the last 200 time slots.}
\label{fig:percentiles}
\end{figure}

To evaluate the proposed LLM-based AI agent, we used a system-level simulator of the UCCF MIMO network proposed by us in~\cite{hoffmann2024}. The simulator utilizes, e.g., a realistic 3D Ray Tracer radio channel model proper for OFDM-based MMIMO systems, a dedicated space-time-frequency RRS, allocation of Modulation and Coding Schemes, {\MH power consumption model~\cite{Hoffmann2021}}, and nonlinear characteristics of the transmitter's radio front end, which generates Rapp-modeled distortion in the Power Amplifier (PA). The simulation parameters are summarized in Table~\ref{tab:parameters}.
\begin{table}[!t]
\centering
\caption{Simulation parameters.\label{tab:parameters}}
\newcolumntype{C}{>{\centering\arraybackslash}X}
\begin{tabularx}{0.45\textwidth}{|CC|}
\hline
\textbf{Parameter}	& \textbf{Value}\\
\hline
Duration of simulation & 1000 time slots of 0.5~ms\\
Center frequency		& 3.6~GHz \\
Number of RBs & 69 \\
Subcarrier spacing & 30~kHz \\
Number of CPU & 1 \\
Number of BSs & 6 - (1 macro, 5 micro) \\
BSs height & macro BS:45~m, micro BS: 6~m micro \\
PA model & Rapp with smoothing factor 12 \\
Maximum Transmit power & macro BS: $46$~dBm, micro BS: $30$~dBm \\
Input Back-Off & 6~dB \\
Number of antennas & macro BS: 128, micro BS: 32 \\
Maximum number of spatial layers & macro BS: 32, micro BS:8 \\
3D Ray Tracer configuration & 15 reflections, 1 diffraction\\
Number of UEs & 40 \\
Urban scenario & Madrid Grid Model \\
\hline
\end{tabularx}
\end{table}
During simulation studies, we compare the proposed LLM-based AI agent under 3 underlying models: \emph{GPT-5}, \emph{GPT-4o}, and \emph{GPT-4o-mini} {\MH (the simulations were done in November 2025 using the latest versions of GPT models)}. In addition, we compare its performance against the \emph{Network Centric} approach (each user is served by a single BS providing the highest received power), \emph{Location-Based} approach~\cite{D’Andrea2020} (the closest 20 BSs serve the user), and the \emph{$\Delta=0.95$} approach~\cite{cfmimobook2021} (the user is served by the minimum number of BSs providing at least 95\% of the total received power). The {\MH resultant EE} over time achieved by each approach is depicted in Fig.~\ref{fig:training}. To smooth the temporal variations related, e.g., to RRS in time, a moving average of 100 slots is applied. The \emph{Network Centric} provides an {\MH EE of around 68 kbit/J}. This serves as a reference value for the state-of-the-art systems. Both \emph{Location-Based}, and \emph{$\Delta=0.95$} approaches are characterized by a degradation of the {\MH EE} reaching the values of about {\MH 50 and 40 kbit/J}, respectively. This is the result of serving the UEs with too many BSs, causing, e.g., that the transmit power is inefficiently allocated to the UE being far away from a particular BS. Moreover, these algorithms do not come with an embedded ability to modify their parameters based on feedback from the network. Thus, they stick to the initially formulated serving cluster, even if it decreases performance compared to the \emph{Network Centric} approach. On the other hand, it can be seen that the EE obtained by the proposed AI agent depends on the LLM model used. The \emph{GPT-4o-mini} is the least advanced one and is characterized by an EE similar to the \emph{Network Centric} approach. For both \emph{GPT-4o} and \emph{GPT-5}, fluctuations can be observed in Fig.~\ref{fig:training}, indicating that the AI agent gains knowledge and explores potential serving clusters for UCCF MMIMO network users. Interestingly, during the last 200 slots, when both models have stabilized solution, the \emph{GPT-4o} achieves an {\MH EE of about 90 kbit/J, outperforming \emph{GPT-5}, which achieves only about 80 kbit/J. This gives up to 32\% gain in EE for \emph{GPT-4o} compared to the \emph{Network Centric} approach.}
To better understand the degrees of difference in solutions of the proposed LLM-based AI agent under different LLM models, in Fig.~\ref{fig:decisions}, we investigated the cumulative number of decisions to modify the serving cluster for a UE over time, for each considered GPT model. By a single decision, we understand the change of a serving cluster size for a particular UE, e.g., if a given UE was served by 1 BS and in the next iteration is served by 3 BSs, there were 2 decisions made. In Fig.~\ref{fig:decisions} we can see that, while using the \emph{GPT-4o-mini}, there were only decisions made at the beginning, and throughout the simulation, the AI agent did not make any adjustments. This explains its poor performance compared to the \emph{GPT-5}, and \emph{GPT-4o}, which made over 35 and over 65 decisions, respectively. We can expect that this disproportion in the decisions made is reflected by the better SCF obtained under \emph{GPT-4o}, which resulted in a higher {\MH EE} compared to \emph{GPT-5}, i.e., \emph{the GPT-4o} model did more exhaustive exploration and adjustment based on the feedback from the UCCF MMIMO network.
Finally, we investigated how the SCF aimed at improving {\MH EE} done by the proposed LLM-based AI agent affects the throughput of the UEs under poor, normal, and very good radio conditions represented as 10th, median, and 90th percentiles from the UE throughput distribution, respectively. The results obtained over the last 200 slots when the algorithms become stable are depicted in Fig.~\ref{fig:percentiles}. It can be seen that, with respect to the 10th percentile compared to the \emph{Network Centric} approach, there is an improvement of about 4\% and 0.5\%, for \emph{GPT-5} and \emph{GPT-4o} respectively. In the case of \emph{GPT-4o-mini}, \emph{Location-Based}, and \emph{$\Delta=0.95$} there is a degradation of 32\%, 10\% and 18\%, respectively. Regarding the median UE throughput, the \emph{GPT-4o} is characterized by the best performance, providing an improvement of about 37\%  compared to the \emph{Network Centric} approach. Regarding the 90th percentile, the \emph{GPT-4o} and \emph{GPT-4o-mini} provide degradation of 1\% compared to the \emph{Network Centric} approach, while the \emph{GPT-5}, \emph{Location-Based} and \emph{$\Delta=0.95$} approaches decrease it by 18\%, 68\% and 71\%, respectively.

\section{Conclusion} \label{sec:conclusions}
We showed that the proposed LLM-based AI agent improves the EE of the UCCF MMIMO network by dynamic SCF by up to 32\%. The simulation study carried out in the realistic environment proved its superiority over the state-of-the-art approaches in terms of EE, while not decreasing the QoS of UEs characterized by the worst and the best radio conditions. Most importantly, the performance of the AI agent depends on the LLM model used. This study showed the highest potential in \emph{GPT-4o}. However, to prove this claim, further studies are required.

\bibliography{references} 
\bibliographystyle{IEEEtran}

\end{document}